\documentclass[twoside,a4paper,11pt]{proceedings}
\usepackage{graphicx}
\usepackage{hyperref}
\usepackage{amsmath}
\usepackage{txfonts}
\usepackage{movie15}
\usepackage{gensymb}
\usepackage{lipsum}
\usepackage{comment}
\usepackage{multirow}
\usepackage{changepage}
\usepackage{float}
\usepackage{appendix}
\usepackage{tabularx}
\usepackage{epsfig}
\usepackage{upgreek}
\usepackage{natbib}
\let\OLDthebibliography\thebibliography
\renewcommand\thebibliography[1]{
  \OLDthebibliography{#1}
  \setlength{\parskip}{0pt}
  \setlength{\itemsep}{0pt}
}
\begin{document}
\pagenumbering{arabic}
\pagestyle{myheadings}
\thispagestyle{empty}
\vspace*{-1cm}
\flushleft\includegraphics[width=3cm,viewport=0 -30 200 -20]{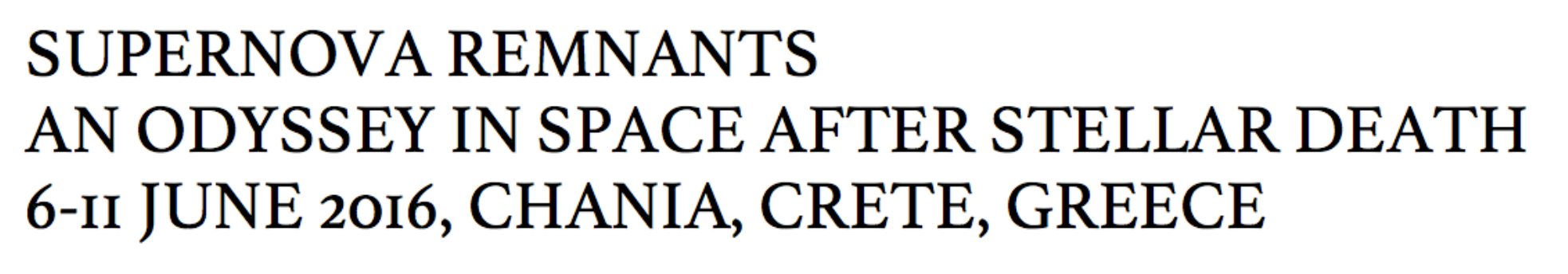}
\vspace*{0.2cm}
\begin{flushleft}
{\bf {\LARGE
The circumstellar environment of pre-SN Ia systems
}\\
\vspace*{1cm}
E. Harvey$^1$,
M. P. Redman$^1$,
P. Boumis$^2$,
M. Kopsacheili$^2$,
S. Akras$^3$,
L. Sabin$^4$
and T. Jurkic$^5$
%
%
}\\
\vspace*{0.5cm}
%
\small
$^{1}$
Centre for Astronomy, School of Physics, National University of Ireland Galway, Ireland \\
$^{2}$
IAASARS National Observatory of Athens, Greece \\
$^{3}$
Observat\'orio de Valongo, Universidade Federale do Rio de Janeiro, Brasil \\
$^{4}$
Instituto de Astronomia, UNAM, Ensenada, Mexico \\
$^{5}$
Department of Physics, University of Rijeka, Croatia
%
\end{flushleft}
\markboth{
Old classical nova shells as CSM around SNIa
}{
E. Harvey et al.
}

\thispagestyle{empty}
\vspace*{0.4cm}
\begin{minipage}[l]{0.09\textwidth}
\ 
\end{minipage}
\begin{minipage}[r]{0.9\textwidth}
\vspace{1cm}
\section*{Abstract}{\footnotesize
Here we explore the possible preexisting circumstellar debris of supernova type Ia systems. Classical, symbiotic and recurrent novae all accrete onto 
roughly solar mass white dwarfs from main sequence or Mira type companions and result in thermonuclear runaways and 
expulsion of the accreted material at high velocity. The expelled material forms a fast moving shell that eventually 
slows to planetary nebula expansion velocities within several hundred years. All such systems are 
recurrent and thousands of shells (each of about 10${^{-4}}$M$_{\odot}$) snow 
plough into the environment. As these systems involve common envelope binaries the material is distributed in a non-spherical shell. These systems could be progenitors of some 
SN Ia and thus explode into environments with large amounts of accumulated gas and dust distributed in 
thin non-spherical shells. Such shells should be observable around 100 years after a SN Ia event in a radio flash 
as the SN Ia debris meets that of the ejected material of the systems previous incarnation.\\
\vspace{10mm}
}
\end{minipage}

\section{Introduction}

$\,\!$\indent Classical novae are exciting objects that are well observed during outburst, but suffer from 
lack of attention during their quiescence. All classical novae are believed to reoccur, however only 
those that are seen to are called recurrent novae and do so on human timescales. The more 
frequent a nova's recurrence the closer it gets to the Chandrasekhar limit. To gain a fuller 
understanding of the character of evolving nova shells, a campaign to study their morphology, structure 
and ionization was undertaken. As there are few ({\raise.17ex\hbox{$\scriptstyle\sim$}}40) known nova shells, a 
search through IR archives 
with optical follow-up, has led to the discovery of additional 
shells. In order to decipher the spatial and velocity constraints of these objects, long-slit high-resolution 
spectroscopy paired with imaging allows for the construction of three-dimensional morpho-kinematic 
models. 

The time evolution of the ionization structure of novae can also be followed using multi-epoch 
archival low and medium resolution spectra, which can be simulated using {\sc Cloudy} \citep{2} and {\sc pyCloudy} \citep{3}. The result is a spatial 
map of specified emission lines and thus modelling the nova shell while quiescent 
allowing for a greater understanding of nova evolution and giving insight towards the environment as output
Nova systems live at the cross-roads of evolution regarding some of the most intensely studied stellar phemonena in 
astronomy. Classical, symbiotic and recurrent novae occur within cataclysmic variable systems containing a white dwarf and a companion. 
Symbiotic novae occur within wide binary systems with a Mira and WD,
 simultaneously pre-PN and post-PN.
Classical novae are thermonuclear runaway events within tight binaries
on the surface of a WD with a MS star companion from which it accretesâ again pre-PN and post-PN.
Symbiotic, classical, recurrent novae and PNe all share morphological characteristics and are possible SN Ia progenitor systems.

\section{Observations}
$\,\!$\indent Considering the detectability of  the nova shell and planetary nebula surrounding the GK Per system it was decided to search through the 
WISE archive \citep{4} for hints of IR nebulosity surrounding systems known to harbour classical novae. To follow was an 
investigation to see if this method would show any results by narrow-band ( H$\alpha$+[N~{\sc ii}] and [O~{\sc iii}]) optical observations with the Aristarchos 2.3m telescope 
of previously undiscovered shells in the optical regime. Also, some brighter previously known nova shells and other 
surrounding material were imaged with the discovery of previously unobserved features in certain systems such as ballistic knots, an ancient PN masquerading as an apparent jet \cite{5} and other nebulousity.


To gain velocity information on the aforementioned imaged features, long-slit echelle 
spectroscopy were obtained with the Manchester Echelle Spectrometer on the 2.12m telescope located at the San Pedro Martir 
observatory. To date we have concentrated on nova shells that are $> 100$ years old, with additional observations of younger nova shells due to take place in August 2016.

\section{Modelling}
The narrow band H$\alpha$+[N~{\sc ii}] Aristarchos imaging coupled with the MES spectroscopy can then be interpreted with the 
3D morpho-kinematic code {\sc shape} \citep{1}. 
By building 3D spatial structures 
with velocity, temperature, density plus other modifiers a comprehensive model of a nova shell can be made, with position-velocity arrays or other moment maps as outputs.


From low and medium resolution spectra photoionisation models can be built in either {\sc Cloudy} or {\sc pyCloudy}. 
Through combination of 3D {\sc shape} models and 1D photoionization models we extract pseudo 3D emission models, see Fig. 1.

\begin{figure}[!ht]
\centering
\includegraphics[width=15cm]{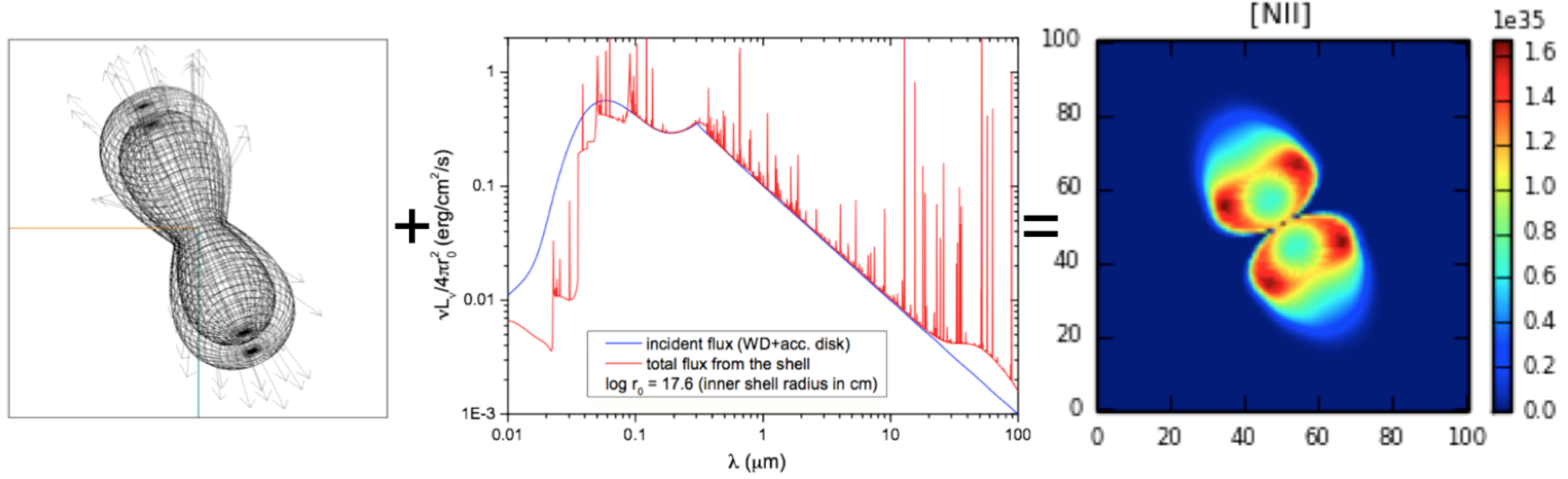}
\caption{\footnotesize From left to right: a  {\sc shape} model of a bipolar nebula with velocity field lines included, predicted {\sc Cloudy}  nova shell spectrum, last two panels are {\sc pyCloudy}  output showing the expected emission from [N~{\sc ii}] 6583 $\AA$.}
\label{fig1}
\end{figure}

\section{Conclusions: Recurrent novae growing towards M$\rm{_{Ch}}$}

$\,\!$\indent 
For the pre-SN Ia systems described here it is possible for their ejected and swept-up matter to accumulate more than 1 M$_{\odot}$ being swept 
up to a distance of roughly 1 pc. However, before undergoing its SN Ia event the system is 
expected to undergo nova explosions every {\raise.17ex\hbox{$\scriptstyle\sim$}}2 months such that the SN Ia ejecta should encounter 
the most recent nova ejecta around day 6 after ejection, although given the density and energy regimes the SN event would completely dominate the single shell interactions. It is only the accumulated $\lq$swept-up' shells that would be detectable with current technology. Since the ejected shells of novae are, in general, 
non-spherical, and more often than not fragmented into discrete clumps, the material into which the SN Ia is expanding is not uniformly distributed. Estimates for when one of these systems may undergo its SN Ia event can be calculated from equations in \cite{6}.


SN Ia occur in complex environments where previous stages of evolution, such as the planetary nebula and
cataclysmic variable stages lead to knots, bipolarity, equatorial and polar over-densities surrounding the objects. Through the collective ejection episodes and swept up ISM several solar masses of debris is expected to accumulate in 
axisymmetric distributions around the SN Ia progenitor systems. This could introduce variations between type Ia SN 
systems dependent on the line-of-sight towards the object, or even produce the $\lq$ears' seen in SN Ia remnants. These 
features can however be modelled in detail during the progenitor stages of evolution.


\scriptsize  
%
\section*{Acknowledgments}   
%
The authors would like to thank the staff at SPM and Helmos observatories for the excellent support received 
during observations. The Aristarchos telescope is operated on Helmos Observatory by the IAASARS of the National Observatory of Athens. 
This publication makes use of data products from the Wide-field Infrared Survey Explorer, which is a joint project of the University of California, Los Angeles, and the Jet Propulsion Laboratory/California Institute of Technology, funded by the National Aeronautics and Space Administration.
E. Harvey wishes to acknowledge the support of the Irish Research Council for providing funding under their postgraduate research scheme. S. Akras gratefully acknowledges a postdoctoral fellowship from the Brazilian Agency CAPES (under their program: "Young Talents Attraction"- Science Without Borders; A035/2013).
The authors greatly benefitted from discussions with T. Jurkic and C. Neilson. 


\end{document}